\documentclass[letterpaper,twocolumn,showpacs,pra,aps]{revtex4-1}

\usepackage{amsmath}  
\usepackage{amsfonts} 
\usepackage{graphicx} 
\usepackage{amssymb}
\usepackage{hyperref}
\usepackage{url}

\begin{document}

\title{Ground-state properties of dilute Bose systems with synthetic dispersion laws}

\author{Eugene B. Kolomeisky}

\affiliation
{Department of Physics, University of Virginia, P. O. Box 400714,
Charlottesville, Virginia 22904-4714, USA}

\date{\today}

\begin{abstract}
Experimental advances in synthesizing spin-orbit couplings in cold atomic Bose gases promise to create single-particle dispersion laws featuring energy minima that are degenerate  on a ring  or a sphere in momentum space.  We show that for arbitrary space dimensionality the ground-state properties of a dilute system of spin-orbit coupled Bose particles with such dispersion and short-range repulsive interactions are universal:  the chemical potential exhibits a quadratic dependence on the particle density as found in a one-dimensional free Fermi gas.          
\end{abstract}

\pacs{67.10.-j, 67.85.-d}

\maketitle

\section{Introduction}

Past decade has witnessed a surge of interest in the physics of systems whose elementary excitations obey dispersion laws that substantially deviate from those traditionally encountered in condensed matter physics.  Since unusual dispersion laws often imply remarkable physical properties, this research activity holds significant applied promise.  Equally important is an improvement of understanding of some of the fundamental physics issues.  For example, the pseudo-relativistic dispersion law for  low energy electrons in graphene links the physics of that system with quantum electrodynamics (QED) and makes it possible to probe otherwise unaccessible regimes of QED parameter values \cite{graphene_review}.   

One of the dispersion laws whose consequences are currently being actively explored features a minimum along a circle (in $d=2$ dimensions) or a sphere (in $d=3$ dimensions) of fixed radius $\hbar q_{0}$ in momentum space so that in its vicinity the excitation energy can be expanded as:
\begin{equation}
\label{Rashba-roton_dispersion law}
\epsilon(\textbf{k})=\frac{\hbar^{2}(k-q_{0})^{2}}{2m}
\end{equation}
where $\textbf{k}$ is the wave vector, $k=|\textbf{k}|$, and $m$ is the effective mass; the zero of the excitation energy is hereafter chosen at its minimum.   A textbook example of Eq.(\ref{Rashba-roton_dispersion law}) is the roton minimum in the excitation spectrum of superfluid $He^{4}$ which gives rise to an excess heat capacity at intermediate temperatures \cite{LL9}.  The dispersion law (\ref{Rashba-roton_dispersion law}) also arises in a variety of electron systems:

(i) Rashba spin-orbit coupling (SOC) \cite{{Rashba_materials}} gives a ring of energy minimum in two-dimensional materials, at the smaller eigenvalue of the Bychkov-Rashba (BR) Hamiltonian \cite{BR}:
\begin{equation}
\label{BR_Hamiltonian}
\widehat{H}_{BR}=\frac{\hbar^{2}}{2m}\left (k^{2}+2q_{0}[\widehat{\boldsymbol{\sigma}}\times \textbf{k}]\boldsymbol\nu +q_{0}^{2}\right )
\end{equation}
where $\widehat{\boldsymbol\sigma}$ stands for the Pauli matrices while $\boldsymbol\nu$ is a unit vector perpendicular to the plane of the electron system.  

(ii) The dispersion law (\ref{Rashba-roton_dispersion law}) is also encountered in a variety of few-layer systems \cite{few_layer_materials} including biased bilayer graphene \cite{bilayer} where the annular character of the band structure (\ref{Rashba-roton_dispersion law}) is due to the electron charge and not spin.  

(iii) In three dimensions Eq.(\ref{Rashba-roton_dispersion law}) is the smaller eigenvalue of the Hamiltonian with $\widehat{\boldsymbol\sigma} \cdot \textbf{k}$ (Weyl) SOC
\begin{equation}
\label{Weyl_Hamiltonian}
\widehat{H}_{W}=\frac{\hbar^{2}}{2m}\left (k^{2}+2q_{0}\widehat{\boldsymbol\sigma} \cdot \textbf{k}+ q_{0}^{2}\right )
\end{equation}
 
Recent experimental breakthroughs in the synthesis of non-Abelian gauge fields by precise control of interactions of ultracold $^{87}Rb$ atoms with light \cite{gauge} has made it possible to realize of a SOC (with $\widehat{\boldsymbol\sigma}$ corresponding to a pseudo-spin $1/2$ degree of freedom) that leads to a dispersion law $\epsilon(\textbf{k})$ with multiple discrete minima \cite{SO_interaction}.  Bypassing the Pauli spin-statistics theorem ($^{87}Rb$ is a Bose particle), these advances  pave a way to engineer bosonic dispersion laws on demand \cite{Galitskii_Spielman13}.  While the bosonic BR Hamiltonian (\ref{BR_Hamiltonian}) proposed in Ref.\cite{Galitski08} has not been realized yet in the laboratory, its implementation seems plausible.  Moreover, a proposal to engineer a bosonic Weyl SOC Hamiltonian (\ref{Weyl_Hamiltonian}) has been put forward \cite{Anderson}.   

The dispersion law (\ref{Rashba-roton_dispersion law}) is interesting because it exhibits a massive degeneracy along $d-1$ dimensional hypersphere $k=q_{0}$.  As a result as $\epsilon\rightarrow 0$ the density of states (DOS) estimated as $q_{0}^{d-1}dk/d\epsilon \propto 1/\sqrt{\epsilon}$ diverges in a one-dimensional fashion.  Thus the excitation (\ref{Rashba-roton_dispersion law}) is expected to behave in a one-dimensional manner even though the real space isotropy is intact.  Indeed,  a one-dimensional character of the two-roton binding in $He^{4}$ is well-known \cite{Ruvalds}; a similar effect has been also found in the two-dimensional case \cite{BR_binding}.  In the many-body context the one-dimensional nature of the dispersion law (\ref{Rashba-roton_dispersion law}) is expected to play a role in determining the character of the ground state in biased bilayer graphene \cite{Stauber};  it is also responsible for the effect of anomalous screening in Rashba electron systems \cite{Rashba_paper}.  

The case of a quartic-in-momentum dispersion law in two dimensions, 
\begin{equation}
\label{quartic_dispresion}
\epsilon(\textbf{k})=\frac{\hbar^{2}k^{4}}{2mQ^{2}},
\end{equation}
where $Q$ is a parameter having dimensionality of the wave vector is also relevant.  The DOS (estimated as $kdk/d\epsilon\propto 1/\sqrt{\epsilon}$) is again one-dimensional.  This dispersion law (\ref{quartic_dispresion}) is also within experimental reach \cite{quartic_theory} either through the techniques of Refs.\cite{gauge,SO_interaction} or by employing the shaken optical lattice scheme \cite{shaken}.   
 
Sedrakyan, Kamenev and Glazman were the first to point out \cite{suggested} that a dilute system of bosons whose kinematics is governed by the BR Hamiltonian (\ref{BR_Hamiltonian}) may be related to the Tonks-Girardeau limit \cite{Tonks,Girardeau} of a one-dimensional interacting Bose-gas.  The goal of this paper is a demonstration that indeed in the dilute limit the ground-state properties of an interacting system of (pseudo)spin-1/2 bosons obeying the dispersion law (\ref{Rashba-roton_dispersion law}) resemble those of a standard ($\epsilon=\hbar^{2}k^{2}/2m$ dispersion law) one-dimensional interacting Bose-gas.  The latter is known to feature the effect of \textit{fermionization} discovered by Girardeau \cite{Girardeau}, the one-to-one correspondence between ground-state properties and excitation spectrum of point hard-core bosons and free fermions.  By exactly solving the problem of bosons with delta-function repulsion, Lieb and Liniger \cite{Lieb} have further shown that fermionization is a property of the dilute limit, i.e. when the particle density $n$ goes to zero.  On the other hand, the physics in the dilute limit is dominated by pair collisions which in three dimensions allowed for successful application of perturbation theory to calculate ground-state properties of a weakly-interacting Bose-gas \cite{LL9}.  The latter however does not exhibit fermionization. 

A unified picture of the ground-state properties of interacting Bose particles with short-range interactions in the dilute limit for general space dimensionality is supplied by a renormalization-group (RG) approach \cite{KS1,Sachdev}.  Specifically, the fermionization effect present for $d<2$ was found to be a property hinging upon the existence of a nontrivial fixed point of a RG transformation.  Here in the problem of the ground-state properties of interacting BR bosons (\ref{BR_Hamiltonian}) we find a similar fermionization effect:  there exists a non-trivial fixed point of the RG transformation which in the dilute limit is responsible for a quadratic dependence of the chemical potential on the particle density ($\mu\propto n^{2}$) as found in a one-dimensional free Fermi gas.  The same conclusion using the same technique was recently reached in the context of the bosons obeying the quartic dispersion law (\ref{quartic_dispresion}) \cite{Wu}.  RG method has been also employed to study the low-energy physics of spinless bosons with BR dispersion law (\ref{Rashba-roton_dispersion law}) in two dimensions \cite{spinless2};  a related three-dimensional problem has been considered in Ref. \cite{spinless3}.

The possibility of fermionization of the BR bosons (\ref{BR_Hamiltonian}) was considered in the past  \cite{Sedrakyan} where it was argued that the ground state has a composite fermion nature with the chemical potential behaving as $\mu\propto n^{3/2}$.   This state however gives a larger energy per particle compared to what is advocated below.  We hasten to mention that only limited version of fermionization is demonstrated here;  whether the energy spectrum of dilute system of bosons obeying the dispersion laws (\ref{Rashba-roton_dispersion law}) or (\ref{quartic_dispresion}) is fermionic (as is the case of Refs.\cite{Girardeau, Lieb}) or not requires a separate investigation.

\section{T-matrix and Renormalization-Group Analysis}

We proceed along the lines of the previous analysis \cite{KS1} of the ground-state properties of regular bosons focusing on the dispersion law (\ref{Rashba-roton_dispersion law}).  First, in the low-energy limit the pseudo-spin degree of freedom of the particle is locked to its momentum:  the BR boson (\ref{BR_Hamiltonian}) is helical, $\widehat{\boldsymbol{\sigma}} \perp \textbf{k}$, while its Weyl cousin (\ref{Weyl_Hamiltonian}) is chiral, $\widehat{\boldsymbol{\sigma}} \parallel \pm\textbf{k}$.  This fact -- which is built into the dispersion law (\ref{Rashba-roton_dispersion law}) -- allows us to focus exclusively on the translational degrees of freedom.  In the limiting case of slowly colliding identical spin-1/2 particles, scattering only takes place for antiparallel spins \cite{LL3}.  Therefore we consider the scattering of two excitations with wave vectors $\textbf{q}_{0}$ and $-\textbf{q}_{0}$ through intermediate states with wave vectors $\textbf{k}$ and $-\textbf{k}$ under action of the two-body interaction $U(r)=u_{0}\delta_{a}^{d}(\textbf{r})$ where $\delta_{a}^{d}(\textbf{r})$ refers to any well-localized function of range $a$ that transforms into a $d$-dimensional $\delta$-function of strength $u_{0}$ as $a\equiv 1/\Lambda\rightarrow 0$;  the range is assumed to satisfy the condition $aq_{0}=q_{0}/\Lambda\ll 1$.  An exact treatment of the scattering requires replacement of the interaction strength $u_{0}$ with a $t$ matrix which satisfies the equation \cite{t-matrix}
\begin{equation}
\label{t-matrix_eq}
t=u_{0}-u_{0}t\int \frac{d^{d}k}{(2\pi)^{d}}\frac{1}{2\epsilon(\textbf{k})}
\end{equation}
whose solution has the form
\begin{equation}
\label{t-matrix_solution}
\frac{1}{t}=\frac{1}{u_{0}}+\int \frac{d^{d}k}{(2\pi)^{d}}\frac{1}{2\epsilon(\textbf{k})}
=\frac{1}{u_{0}}+\frac{mK_{d}}{\hbar^{2}}\int_{0}^{\Lambda}\frac{k^{d-1}dk}{(k-q_{0})^{2}}
\end{equation}
where $K_{d}$ is the surface area of a $d$-dimensional unit sphere divided by $(2\pi)^{d}$;  the upper integration limit $\Lambda$ is set by the short-ranged behavior of the potential.  

\subsection{Bose system with short-range interactions}

In order to provide a broader context for comparison of our results with what is known, we begin by outlining ground-state properties of the standard Bose system which is the $q_{0}=0$ case of the dispersion law (\ref{Rashba-roton_dispersion law}).  Then for  $d>2$ the integral in (\ref{t-matrix_solution}) converges, $t$ is non-zero, and to leading order as $n\rightarrow 0$ the chemical potential is given by the mean-field (Hartree) expression $\mu=nt$.  On the other hand, for $d\leqslant 2$ the integral in (\ref{t-matrix_solution}) diverges, and the $t$ matrix vanishes which is an indicator of a failure of the mean-field analysis of the many-body problem.  

\subsubsection{Heuristic argument}

This failure can be remedied heuristically by noting that Eq.(\ref{t-matrix_solution}) describes the renormalization of the two-body interaction due to zero-point fluctuations of all wave vectors up to $\Lambda$.  In the many-body  case this renormalization is suppressed for the wave vectors below some typical value of $k^{*}$ at which the chemical potential is comparable to the kinetic energy for that wave vector, $\mu\simeq\hbar^{2}k^{*2}/m$ \cite{Popov,FH}.  Thus the many-body nature of the problem effectively imposes a finite lower integration limit in Eq.(\ref{t-matrix_solution}) so that for $d\leqslant 2$ the $t$ matrix remains nonzero, acquiring a dependence on the chemical potential according to
\begin{equation}
\label{heuristic_solution_conventional_bosons}
\frac{1}{t(\mu)}\approx\frac{1}{u_{0}}+\frac{mK_{d}}{\hbar^{2}}\int_{\simeq\frac{\sqrt{m\mu}}{\hbar}}^{\Lambda}k^{d-3}dk
\end{equation} 
The solution to the two-body body scattering problem (\ref{t-matrix_solution}) is relevant to the many-body case when the upper and lower integration limits in (\ref{heuristic_solution_conventional_bosons}) are well-separated ($\mu\ll\hbar^{2}\Lambda^{2}/m$), which is a condition of the dilute limit adopted hereafter.  

For $d=2$ the precise values of the integration limits in (\ref{heuristic_solution_conventional_bosons}) are unimportant and one finds with logarithmic accuracy $t(\mu)\approx 4\pi \hbar^{2}/m\ln(\hbar^{2}\Lambda^{2}/m\mu)$.  Combining this with the modified Hartree condition $\mu=nt(\mu)$ and solving for the chemical potential, one obtains $\mu\approx 4\pi \hbar^{2}n/m\ln(1/na^{2})$ which is a well-known result \cite{Schick,Popov}.  Its hallmark is near-universality:  the bare interaction strength $u_{0}$ drops out (entering only through the range of applicability of the result) while the dependence on the microscopic length scale $a\equiv1/\Lambda$ is logarithmically weak.      

For $d<2$ one similarly finds $t(\mu)\simeq(\hbar^{2}/m)(m\mu/\hbar^{2})^{(2-d)/2}$;  combining this with the Hartree condition $\mu=nt(\mu)$ recovers the universal result \cite{KS1}
\begin{equation}
\label{ff_general_d}
\mu\simeq\frac{\hbar^{2}n^{2/d}}{m}
\end{equation}
formally coinciding with an expression for the chemical potential of a $d$-dimensional free-fermion gas.  

\subsubsection{Renormalization-group equations}

While the heuristic argument captures the physics of the problem highlighting the interplay of zero-point fluctuations and many-body effects, the RG treatment explains the origin of these conclusions.  The RG equations for $q_{0}=0$ can be derived via a repeated partial integration in Eq.(\ref{t-matrix_solution}) over infinitely narrow $[\Lambda(1-dl); \Lambda]$ slice of the wave vector range followed by a scaling transformation which restores Eq.(\ref{t-matrix_solution}) to its original form with renormalized $u(l)$ obeying the \textit{exact} equation \cite{FH,KS1,KS2,Sachdev}
\begin{equation}
\label{RG_equation_interaction}
\frac{du(l)}{dl}=(2-d)u(l)-\frac{mK_{d}\Lambda^{d-2}}{\hbar^{2}}u^{2}(l),~~~~~u(0)=u_{0}
\end{equation} 
For attractive interactions, $u_{0}<0$, this equation describes the two-body binding problem \cite{KS2} while for repulsive interactions any "initial"  $u_{0}>0$ "flows" as $l\rightarrow \infty$ toward the trivial, $u=0$ ($d>2$), or nontrivial, $u^{\star}\simeq\hbar^{2}\Lambda^{2-d}/m$ ($d<2$), fixed points of (\ref{RG_equation_interaction});  the fixed points coalesce in the marginal $d=2$ case.  The physical meaning of the fixed points becomes clear in the many-body problem when Eq.(\ref{RG_equation_interaction}) is supplemented by two additional equations \cite{FH} describing the renormalization of the chemical potential $\mu(l)$ and the particle density $n(l)$
\begin{equation}
\label{chempotential_density}
\mu(l)=\mu e^{2l},~~~n(l)=ne^{dl}
\end{equation}
which follow from dimensional considerations.  The ground-state properties can be extracted from the Hartree relationship between the renormalized quantities \cite{FH}
\begin{equation}
\label{renormalized_Hartree}
\mu(l)=n(l)u(l)
\end{equation} 
When the expression for $\mu(l)$ in Eq.(\ref{chempotential_density}) is substituted (instead of $\mu$) into the condition of the dilute limit ($\mu \ll \hbar^{2}\Lambda^{2}/m$), the latter becomes invalid on a scale 
\begin{equation}
\label{interruption_scale}
l^{*}\simeq \ln\frac{\hbar\Lambda}{\sqrt{m\mu}}\gg 1
\end{equation}
This corresponds to the wave vector $k^{*}\simeq \Lambda e^{-l^{*}}\simeq\sqrt{m \mu}/\hbar$ that already appeared as the lower integration limit in Eq.(\ref{heuristic_solution_conventional_bosons});  the RG flow is interrupted on the scale $l^{*}$.  This scheme provides a comprehensive picture of the ground-state properties of dilute Bose systems \cite{KS1} for general $d$;  specifically, the fermionization present for $d<2$ is due to the flow toward the nontrivial (free-fermion) fixed point $u^{\star}\simeq \hbar^{2}\Lambda^{2-d}/m$ of Eq.(\ref{RG_equation_interaction}).  This is easy to see because for $l=l^{*}$,  Eq.(\ref{renormalized_Hartree}) becomes $\mu(l^{*})\approx n(l^{*})u^{\star}\simeq n(l^{*})\hbar^{2}\Lambda^{2-d}/m$.  Substituting here the expressions for $\mu(l^{*})$, $n(l^{*})$, and $l^{*}$ from Eqs.(\ref{chempotential_density}) and (\ref{interruption_scale}) recovers Eq.(\ref{ff_general_d}).  Since the interruption scale (\ref{interruption_scale}) is an order of magnitude estimate, the RG treatment cannot recover a numerical factor missing from Eq.(\ref{ff_general_d}).

\subsection{Spin-orbit coupled system of bosons with short-range interactions}

For $q_{0}$ finite, the integral in Eq.(\ref{t-matrix_solution}) diverges regardless of the space dimensionality which means that the $t$ matrix is zero.  

\subsubsection{Heuristic argument}

This resembles the situation in a conventional Bose system for $d\leqslant 2$, and the outcome can be understood via a heuristic argument similar to the one which led to Eq.(\ref{heuristic_solution_conventional_bosons});  now the $t$ matrix depends on the chemical potential according to
\begin{equation}
\label{heuristic_solution_Rashba_bosons}
\frac{1}{t(\mu)}\approx \frac{1}{u_{0}}+\frac{2mK_{d}q_{0}^{d-1}}{\hbar^{2}}\int_{\simeq\sqrt{m\mu}/\hbar}^{\Lambda}\frac{dk'}{k'^{2}}
\end{equation}
where $\sqrt{m\mu}/\hbar$ corresponds to the typical width of the hyperspherical layer centered around $k=q_{0}$ wherein the many-body effects suppress the downward renormalization of the two-body interaction.  Incidentally, Eq.(\ref{heuristic_solution_Rashba_bosons}) has the same form as the $d=1$ case of Eq.(\ref{heuristic_solution_conventional_bosons}) describing conventional bosons.   Computing the integral and combining the outcome with the condition $\mu=nt(\mu)$ leads to our central result
\begin{equation}
\label{Rashba_ff_1d}
\mu\simeq \frac{\hbar^{2}n^{2}}{mq_{0}^{2(d-1)}},~~~~~~~~\frac{n}{q_{0}^{d-1}\Lambda}\ll1, \frac{\hbar^{2}n}{mu_{0}q_{0}^{2(d-1)}}\ll1
\end{equation}   
This expression for the chemical potential coincides with that of a $d$-dimensional free-fermion gas of particles obeying the dispersion law (\ref{Rashba-roton_dispersion law}).  At the same time, the quadratic dependence on the particle density ($\mu\propto n^{2}$ -- compare this to Eq.(\ref{ff_general_d}) for $d=1$) is a hallmark of the pseudo-one-dimensional character of the result (\ref{Rashba_ff_1d}).  As its range of applicability indicates, the conclusion holds in the dilute limit $n\rightarrow 0$.  At the same time, for $n$ fixed and \textit{point} interactions ($a=1/\Lambda=0$) the inequality $n/q_{0}^{d-1}\Lambda\ll1$ holds automatically, and we are  left with only the second $\hbar^{2}n/mu_{0}q_{0}^{2(d-1)}\ll1$ constraint which parallels the condition of the dilute limit of Lieb and Liniger \cite{Lieb} found for conventional bosons in a strictly one-dimensional case.  Taking further the \textit{hard-core} $u_{0}=\infty$ limit automatically satisfies the remaining $\hbar^{2}n/mu_{0}q_{0}^{2(d-1)}\ll1$ condition.  Therefore for point hard-core bosons the result (\ref{Rashba_ff_1d}) is expected to be exact (no dilute corrections) which parallels Girardeau's result \cite{Girardeau} in the strictly one-dimensional case.  We note that due to violation of the inequalities in Eq.(\ref{Rashba_ff_1d}) the $q_{0}\rightarrow 0$ limit cannot be taken;  this is a consequence of asymptotic character of Eq.(\ref{heuristic_solution_Rashba_bosons}) that only accounts for the leading divergence in Eq.(\ref{t-matrix_solution}).

\subsubsection{Renormalization-group equations}

These conclusions can be put on a solid footing and a connection to fermionization of conventional bosons can be made clear by use of a RG method.  First, we split the integration range in Eq.(\ref{t-matrix_solution}) into two segments, $[0;\Lambda(1-dl)]$ and $[\Lambda(1-dl);\Lambda]$, and carry out a partial integration over the latter, finding
\begin{equation}
\label{partial_integration}
\frac{1}{t}\approx \frac{1}{u_{0}}+\frac{2mK_{d}q_{0}^{d-1}}{\hbar^{2}\Lambda}dl+\frac{mK_{d}}{\hbar^{2}}\int_{0}^{\Lambda(1-dl)}\frac{k^{d-1}dk}{(k-q_{0})^{2}}
\end{equation}          
where, like in Eq.(\ref{heuristic_solution_Rashba_bosons}), the second term in the right-hand side is written in an approximation that captures the leading divergence in (\ref{t-matrix_solution}).  Changing the variable in the integral to $k= k'(1-dl)$ and dividing both sides by $(1-dl)^{d-2}$ restores the original form of Eq.(\ref{t-matrix_solution}) except that $t$, $u_{0}$, and $q_{0}$ are replaced with their renormalized counterparts $t(l)$, $u(l)$, and $q(l)$;  the last two obeying differential equations
\begin{equation}
\label{ RG_Rashba_equation_interaction}
\frac{du(l)}{dl}=(2-d)u(l)-\frac{2mK_{d}}{\hbar^{2}\Lambda}u^{2}(l)q^{d-1}(l)
\end{equation}
\begin{equation}
\label{RG_Rashba_radius}
\frac{dq(l)}{dl}=q,~~~q(0)=q_{0}
\end{equation}
We note that the first terms in the right-hand sides of Eqs.(\ref{RG_equation_interaction}) and (\ref{ RG_Rashba_equation_interaction}) are the same because they reflect identical scaling transformation of the interaction.  Similarly, Eq.(\ref{RG_Rashba_radius}) reflects the scaling transformation of the wave vector.  Increase of $q(l)$ under rescaling is a sign that it is a perturbation relevant in the RG sense.  Eqs.(\ref{ RG_Rashba_equation_interaction}) and (\ref{RG_Rashba_radius}) replacing Eq.(\ref{RG_equation_interaction}) is the only change in the general approach needed to understand the ground-state properties of the BR bosons.  Introducing dimensionless interaction strength
\begin{equation}
\label{dimensionless_interaction}
v(l)=\frac{2mK_{d}}{\hbar^{2}\Lambda}q^{d-1}(l)u(l)
\end{equation}
reduces Eqs.(\ref{ RG_Rashba_equation_interaction}) and (\ref{RG_Rashba_radius}) to a single equation
\begin{equation}
\label{RG_equation_dimensionless_interaction}
\frac{dv(l)}{dl}=v(l)-v^{2}(l), ~~~v(0)=v_{0}=\frac{2mK_{d}}{\hbar^{2}\Lambda}q_{0}^{d-1}u_{0}
\end{equation} 
which has exactly the same form as the $d=1$ version of Eq.(\ref{RG_equation_interaction}).  Its solution is
\begin{equation}
\label{solution_dimensionless_interaction}
v(l)=\frac{1}{1-(1-v_{0}^{-1})e^{-l}}
\end{equation}
For a repulsive interaction any initial $v_{0}>0$ flows toward the stable nontrivial fixed point $v^{\star}=1$ which is the reason underlying the free-fermion appearance of result (\ref{Rashba_ff_1d}).  Indeed, combining Eqs.(\ref{chempotential_density}), (\ref{dimensionless_interaction}), and solution to Eq.(\ref{RG_Rashba_radius}), $q(l)=q_{0}e^{l}$ evaluated at the interruption scale (\ref{interruption_scale}), one recovers Eq.(\ref{Rashba_ff_1d}). 

\subsubsection{Connection to two-body binding problem}

As an illustration of generality of our analysis we note that for attractive interactions any initial $v_{0}<0$ flows away from the unstable fixed point $v=0$ according to Eq.(\ref{solution_dimensionless_interaction}) and diverges at a finite scale $l_{b}$ which, for weak attraction ($|v_{0}|\ll1$), is $l_{b}=\ln(1/|v_{0}|)$.  According to Ref.\cite{KS2} this is a sign of a two-body bound state with a localization length $\xi=ae^{l_{b}}$ and a binding energy $E_{0}\simeq -\hbar^{2}/m\xi^{2}$ given by
\begin{equation}
\label{2_body_bound_state_properties}
\xi\simeq \frac{\hbar^{2}}{mq_{0}^{d-1}}\frac{1}{|u_{0}|},~~~~~~~~ E_{0}\simeq -\frac{mq_{0}^{2(d-1)}}{\hbar^{2}}u_{0}^{2}
\end{equation} 
The $1/|u_{0}|$ divergence of the localization length and the vanishing of the binding energy according to  $-u_{0}^{2}$ are indicators of the one-dimensional character of the binding which is due to the dispersion law ({\ref{Rashba-roton_dispersion law}).  Eqs.(\ref{2_body_bound_state_properties}) agree with $s$-state binding properties of rotons ($d=3$) \cite{Ruvalds} and BR particles ($d=2$) \cite{BR_binding}.  

\subsubsection{System of bosons with quartic dispersion law}

Having explained both the physics and formalism underlying fermionized form of the ground-state properties  of dilute Bose systems with BR dispersion law (\ref{Rashba-roton_dispersion law}) makes it straightforward to address the problem of the ground-state properties of bosons obeying the quartic (\ref{quartic_dispresion}) (or arbitrary power) dispersion law.  An analysis that closely mirrors the treatment of the standard bosons \cite{KS1} (also outlined earlier in the text) then shows that in the quartic case ({\ref{quartic_dispresion}) the chemical potential (including conditions of the dilute limit) will be given by the $d=2$ version of Eqs.(\ref{Rashba_ff_1d}) with $q_{0}$ replaced by the parameter $Q$ entering the dispersion law (\ref{quartic_dispresion}) thus confirming the result of Ref.\cite{Wu}.  The same replacements need to be made in Eqs.(\ref{2_body_bound_state_properties}) that now will describe two "quartic" bosons bound by weak short-range attractive interaction.  

Finally, our result $\mu\propto n^{2}$ implies that the long-wavelength low energy statics and dynamics of dilute bosons exhibiting the dispersion laws (\ref{Rashba-roton_dispersion law}) or (\ref{quartic_dispresion}) placed in external potentials will be correctly described by a version of the Gross-Pitaevskii theory tailored to the free-fermion limit of one-dimensinal bosons \cite{GP_modified}.   

The author thanks T. A. Sedrakyan for a discussion that stimulated this work, critical comments and literature references and J.P. Straley for comments.

\end{document}